\documentclass[10pt,journal]{IEEEtran}

\usepackage{amsmath,amssymb,amsfonts}
\usepackage{graphicx}
\usepackage{textcomp}
\usepackage{booktabs}
\usepackage{tabularx}
\usepackage{multirow}
\usepackage{amsthm}

\usepackage{float}
\usepackage[caption=false]{subfig}
\usepackage{wrapfig}

\usepackage[mincitenames=1,doi=false,isbn=false,url=false,sorting=none]{biblatex} % 
\begin{document}

\title{Rethinking Non-idealities in Memristive Crossbars for Adversarial Robustness in Neural Networks}
\author{Abhiroop Bhattacharjee, and Priyadarshini Panda\\ Department of Electrical Engineering, Yale University, USA}

\date{April 2021}

\maketitle

\begin{abstract}
	\label{abstract}
    \textit{Deep Neural Networks} (DNNs) have been shown to be prone to adversarial attacks. Memristive crossbars, being able to perform \textit{Matrix-Vector-Multiplications} (MVMs) efficiently, are used to realize DNNs on hardware. However, crossbar non-idealities have always been devalued since they cause errors in performing MVMs, leading to computational accuracy losses in DNNs. Several software-based defenses have been proposed to make DNNs adversarially robust. However, no previous work has demonstrated the advantage conferred by the crossbar non-idealities in unleashing adversarial robustness. We show that the intrinsic hardware non-idealities yield adversarial robustness to the mapped DNNs without any additional optimization. We evaluate the adversarial resilience of state-of-the-art DNNs (VGG8 \& VGG16 networks) using benchmark datasets (CIFAR-10, CIFAR-100 \& Tiny Imagenet) across various crossbar sizes. We find that crossbar non-idealities unleash significantly greater adversarial robustness ($>10-20\%$) in crossbar-mapped DNNs than baseline software DNNs. We further assess the performance of our approach with other state-of-the-art efficiency-driven adversarial defenses and find that our approach performs significantly well in terms of reducing adversarial loss.
\end{abstract}

\begin{IEEEkeywords}
Deep Neural Networks, Memristive crossbars, Non-idealities, Adversarial robustness
\end{IEEEkeywords}

\section{Introduction}
\label{sec:introduction}
In the recent years, memristive crossbar systems have received significant focus for their ability to realize \textit{Deep Neural Networks} (DNNs) by efficiently computing analog dot-products~\cite{schuman}. These systems have been realized using a wide range of emerging technologies such as, \textit{Resistive Random Access Memory} (ReRAM), \textit{Phase Change Memory} (PCM) and Spintronic devices~\cite{chakraborty2020pathways}. These devices exhibit high on-chip storage density, non-volatility, low leakage and low-voltage operation and thus, enable compact and energy-efficient implementation of DNNs~\cite{chakraborty2020pathways, rxnn}. 

Despite so many advantages, the analog nature of computation of dot-products in crossbars poses certain challenges owing to device-level and circuit-level non-idealities such as, interconnect parasitics, process variations in the synaptic devices, driver and sensing resistances, etc.~\cite{geniex,rxnn}. Such non-idealities lead to errors in the analog dot-product computations in the crossbars, thereby adversely affecting DNN implementation in the form of accuracy degradation. Numerous frameworks have been developed in the past to model non-idealities present in crossbar systems and accordingly, retraining the weights (stored in synaptic devices) of the DNNs to mitigate accuracy degradation~\cite{xchangr, indranil, geniex, liu}. \textit{GenieX}~\cite{geniex} is a recent work that uses a neural network-based approach to accurately encapsulate the effects of both data dependent and non-data dependent non-idealities in crossbars and assess their impact on computational accuracy loss.

Crossbar non-idealities, thus, have so far been devalued because they lead to accuracy degradation in DNNs. However, an interesting aspect of these non-idealities in providing resilience to DNNs against adversarial attacks has been unexplored. DNNs have been shown to be adversarially vulnerable~\cite{pgd}. A DNN can easily be fooled by applying structured, yet, small perturbations on the input, leading to high confidence misclassification of the input. This vulnerability severely limits the deployment and potential safe-use of DNNs for real world applications such as self-driving cars, malware detection, healthcare monitoring systems etc.~\cite{carlini, quanos}. Thus, it is imperative to ensure that the DNN models used for such applications are robust against adversarial attacks. Recent works such as~\cite{pixeld, defquant} show quantization methods, that primarily reduce compute resource requirements of DNNs, act as a straightforward way of improving the robustness of DNNs against adversarial attacks. In \cite{pixeld, quanos}, the authors show that efficiency-driven hardware optimization techniques can be leveraged to improve adversarial resilience, while yielding energy-efficiency. In this work, we present a comprehensive analysis on how device-level and circuit-level non-idealities intrinsic to analog crossbars can be leveraged for adversarial robustness in neural networks. To the best of our knowledge, we are the first to show that the intrinsic hardware variations manifested through non-idealities in crossbars intrinsically improve adversarial security without any additional optimization. Our main finding is that- \textit{A DNN model mapped on hardware, while suffering accuracy degradation, is also more adversarially resilient than the baseline software DNN}. 

\textbf{Contributions:} In summary, the key contributions of this work are as follows:
\begin{itemize}
    \item  We develop a systematic framework in \textit{Pytorch} to map DNNs onto memristive crossbar arrays and investigate the cumulative impact of various circuit and device-level non-idealities to confer adversarial robustness. 
    
    \item We analyse the robustness of state-of-the-art DNNs, \textit{viz.} VGG8 and VGG16 using benchmark datasets- CIFAR-10, CIFAR-100 and Tiny Imagenet across various crossbar dimensions. 
   
    \item We find that non-idealities lead to higher adversarial robustness ($>10-20\%$) in DNNs mapped onto crossbars than baseline software DNNs.
    
    \item We investigate the role of various crossbar parameters (such as, $R_{MIN}$) in unleashing adversarial robustness to DNNs mapped onto crossbars. We further study the impact of combining various efficiency-driven hardware optimization techniques with hardware non-idealities on the resiliency of neural networks. 
    
    \item A comparison of our proposed method with other state-of-the-art quantization techniques is also presented to emphasise the importance of hardware non-idealities in confering robustness to DNNs against adversarial inputs. 
\end{itemize}

\vspace{-3mm}
\section{Background \& Related Work}
\label{sec:background}
\subsection{Adversarial Attacks and Defenses}
\label{adv}

DNNs are vulnerable to adversarial attacks in which the model gets fooled by applying precisely calculated small perturbations on the input, leading to high confidence misclassification~\cite{quanos}. The authors in~\cite{goodfellow} introduced \textit{Fast Gradient Sign Method} (FGSM) to generate adversarial input ($X_{adv}$) by linearization of the loss function ($L$) of the trained models with respect to the input ($X$):
\begin{equation} \label{eq:Xadv}
X_{adv}~=~X~+~\epsilon~\times~sign(\nabla_{x} L(\theta,X,y_{true}))
\end{equation}
Here, $y_{true}$ is the true class label for the input X; $\theta$ denotes
the model parameters (weights, biases etc.) and $\epsilon$ quantifies
the degree of distortion. 
The quantity $\Delta=\epsilon~\times~sign(\nabla_{x} L(\theta,X,y_{true}))$ is the net perturbation added to the input ($X$), which is controlled by $\epsilon$. It is noteworthy that gradient propagation is, thus, a crucial step in unleashing an adversarial attack. Furthermore, the contribution of gradient to $\Delta$ varies for different layers of the network~\cite{quanos}. In addition to FGSM-based attacks, multi-step variants of FGSM, such as \textit{Projected Gradient Descent} (PGD)~\cite{pgd} have also been proposed that cast stronger attacks. 

To build resilience against against small adversarial perturbations, defense mechanisms such as gradient masking or obfuscation~\cite{gradmask} have been proposed. Such methods construct a model devoid of useful gradients, thereby making it difficult to create an adversarial attack. In the recent years, several heuristic adversarial defense strategies have been developed, including adversarial training, randomization-based techniques and denoising methods~\cite{pgd, reviewadv}. However, these defenses might be broken by a new attack in the future since they lack a theoretical error-rate guarantee. Hence, researchers have strived to develop certified defensive methods, which always maintain a certain accuracy under a well-defined class of attacks. Even though the certified defense methods indicate a way to reach theoretically guaranteed security, their accuracy and efficiency are far from meeting the practical requirements~\cite{reviewadv}. Apart from these, several quantization-based methods on software have been proposed of late, including works like~\cite{pixeld, defquant, quanos} to improve resilience of neural networks against adversarial perturbations. The work in~\cite{pixeld} deals with discretization of the input space (or allowed pixel levels from 256 values or 8-bit to 4-bit, 2-bit). It shows that input pixel discretization improves the adversarial robustness of DNNs for a substantial range of perturbations, besides improvement in its computational efficiency with minimal loss in test accuracy. Likewise, \textit{QUANOS}~\cite{quanos} is a framework that performs layer-specific hybrid quantization of DNNs based on a metric termed as \textit{Adversarial Noise Sensitivity} (ANS) to make DNNs robust against adversarial perturbations. In contrast to prior works, we present a first of its kind work that comprehensively studies the inherent advantage of hardware non-idealities towards imparting adversarial robustness to DNNs without relying upon other software-based optimization methodologies.

%\textbf{Types of Attacks:}  Broadly, attacks to evaluate adversarial robustness are classified as: \textit{Black-Box} (BB) and \textit{White-Box} (WB). WB attacks are launched when the attacker has complete knowledge of the target model parameters and training information. BB attacks, on the other hand, are launched when the attacker has no knowledge about the target model parameters. %Resilience against WB adversaries also guarantees resilience against the BB ones for similar perturbation ($\epsilon$) range~\cite{quanos}. Thus, all our subsequent experiments are based on WB adversaries for the assessment of adversarial robustness. 

Note, in this work, \textit{Clean Accuracy} ($CA$) refers to the accuracy of a DNN when presented with the test dataset in absence of an adversarial attack. We define \textit{Adversarial Accuracy} ($AA$) as the accuracy of a DNN on the adversarial dataset created using the test data for a given task. \textit{Adversarial Loss} ($AL$) is defined as the difference between $CA$ and $AA$, \textit{i.e}., $AL = CA - AA$. A neural network upon mapping onto hardware would unquestionably suffer computational accuracy loss leading to decrease in $CA$. However, if the value of $AL$ remains smaller for the network on hardware than the software baseline, it implies that the DNN on hardware is more resilient to adversarial attacks. 
\vspace{-3mm}
\subsection{Memristive crossbars and their non-idealities}
\label{nonideal}

\begin{figure}[t]
    \centering
    \subfloat[]{
    \includegraphics[width=0.55\linewidth]{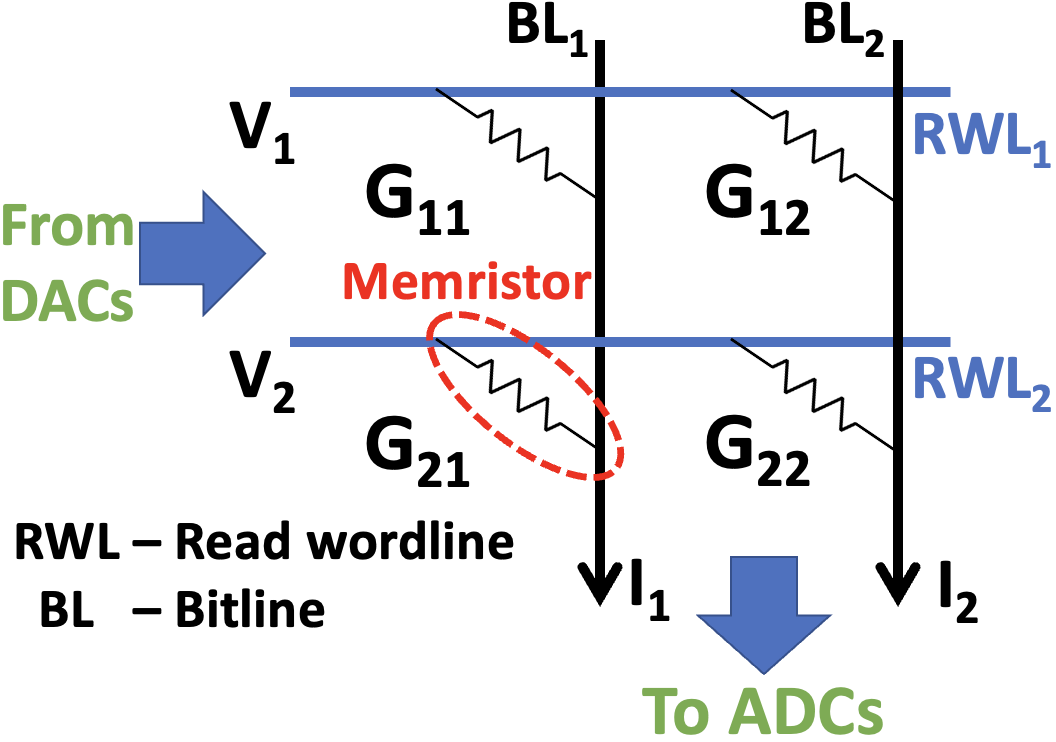}%
    }
    %\hspace{27mm}
    \subfloat[]{
    \includegraphics[width=0.42\linewidth]{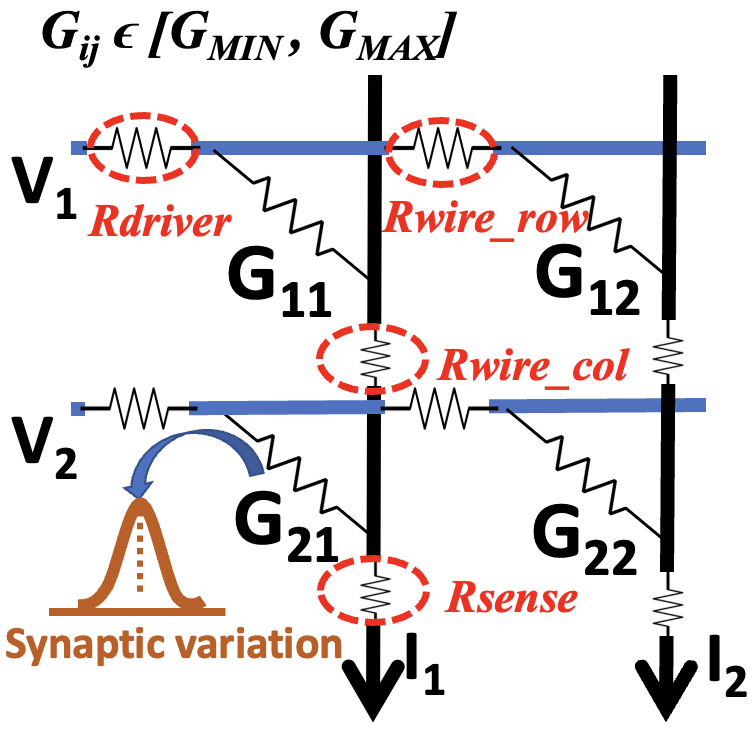}
    }
    %\captionsetup{justification=centering}
    \caption{(a) An ideal 2$\times$2 memristive crossbar array; (b) A typical non-ideal 2$\times$2 crossbar with resistive and device-level non-idealities marked}
    \label{xbar}
    \vspace{-5mm}
\end{figure}

Memristive crossbars have been employed to implement \textit{Matrix-Vector-Multiplications} (MVMs) in an analog manner. Crossbars consist of 2D arrays of synaptic devices, \textit{Digital-to-Analog Converters} (DAC), \textit{Analog-to-Digital Converters} (ADC) and a write circuit. The synaptic devices at the cross-points are programmed to a particular value of conductance (between $G_{MIN}$ and $G_{MAX}$) for a DNN. The MVM operations are performed by converting the digital inputs to the DNN into analog voltages on the Read Wordlines (RWLs) using DACs, and sensing the output current flowing through the Bitlines (BLs) using the ADCs~\cite{rxnn, geniex}. In other words, the activations of the DNNs are mapped as analog voltages $V_i$ input to each row and weights are programmed as synaptic device conductances ($G_{ij}$) at the cross-points as shown in \figurename{~\ref{xbar}}. For an ideal N$\times$N crossbar array (a representative 2$\times$2 ideal crossbar is shown in \figurename{~\ref{xbar}}(a)), during inference, the voltages interact with the device conductances and produce a current (governed by Ohm's Law). Consequently, by Kirchhoff's current law, the net output current sensed at each column $j$ by the ADCs is the sum of currents through each device, \textit{i.e.} $I_{j(ideal)} = \Sigma_{i=1}^{N}{G_{ij} * V_i}$. We term the matrix $G_{ideal}$ as the collection of all $G_{ij}$s for a crossbar instance. %However, in reality, the analog nature of the computation leads to various data-independent (\textit{i.e.}, independent of $V_i$ values) non-idealities, such as, circuit-level resistive non-idealities and device-level variations~\cite{rxnn, geniex}. 

\textbf{Non-idealities:} The analog nature of the computation leads to various non-idealities resulting in errors in the MVMs. These include device-level and circuit-level non-idealities in the memristive crossbars. \figurename{~\ref{xbar}}(b) shows the equivalent circuit for a 2$\times$2 crossbar array highlighting various non-idealities. The circuit-level non-idealities ($Rdriver$, $Rwire\_row$, $Rwire\_col$, $Rsense$) have been modelled as parasitic resistances. The variations in synaptic device conductances owing to the stochasticity of the memristive devices is modelled using a Gaussian distribution around the nominal device conductance ($G_{ij}$). This results in a $G_{non-ideal}$ matrix, with each element $G_{ij}'$ incorporating the effect due to the non-idealities, obtained using circuit laws (Kirchoff's laws and Ohm's law) and linear algebraic operations~\cite{rxnn}. Consequently, for an N$\times$N non-ideal crossbar, the net output current sensed at each column $j$ becomes $I_{j(non-ideal)} = \Sigma_{i=1}^{N}{G_{ij}' * V_i}$, which deviates from its ideal value. This manifests as accuracy degradation for DNNs mapped onto crossbars. The relative deviation of $I_{non-ideal}$ from its ideal value is measured using \textit{Non-ideality Factor} (NF)~\cite{geniex} as: $NF~=~(I_{ideal}-I_{non-ideal})/I_{ideal}$.
%\begin{equation} \label{eq:nf}
%NF~=~(I_{ideal}-I_{non-ideal})/I_{ideal}.
%\end{equation}
Thus, increased non-idealities in crossbars can induce a greater value of NF. This can lead to a significant impact on the computational accuracy of crossbars and therefore, degradation in the accuracy of the DNNs implemented on hardware~\cite{rxnn, geniex}. Note, the aforementioned non-idealities are data-independent, \textit{i.e.} their impact is not determined by the values of analog voltages input to the crossbar~\cite{geniex}.
\vspace{-3mm}

\section{Can non-idealities be leveraged for adversarial robustness?}

\begin{figure}[h!]
\centering
\includegraphics[width=0.4\textwidth]{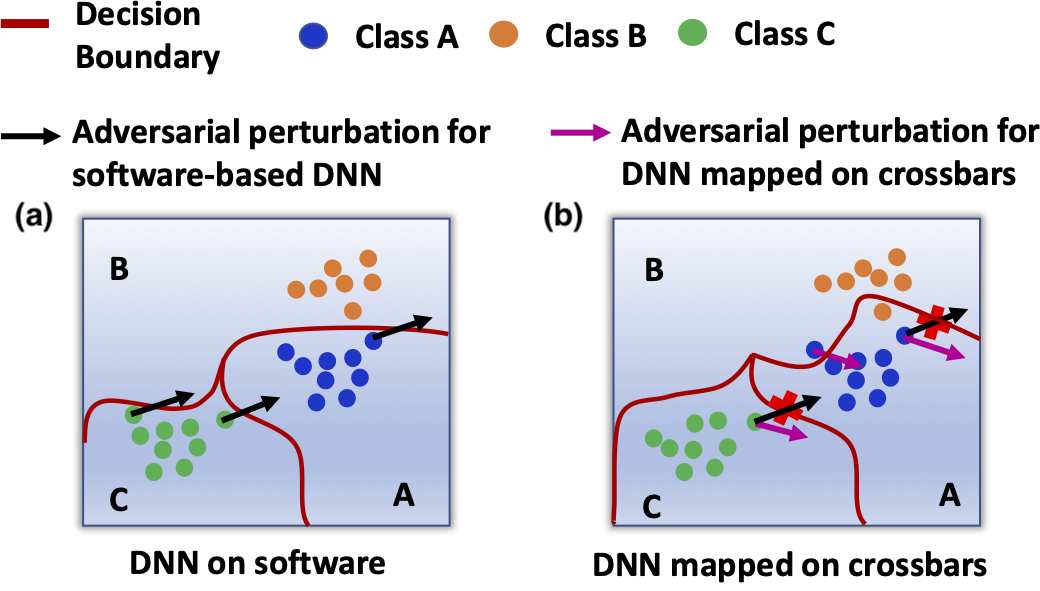}
  
\caption{Pictorial depiction of creation of adversaries for software and hardware-based DNNs - (a) The data points (shown as `dots’) encompass
the data manifold in the high-dimensional subspace. The classifier is trained to separate the data into different classes or hypervolumes based on which the decision boundary is formed. Adversaries are created by perturbing the data points ('black arrow') into a different hypervolume from its typical hypervolume leading to missclassifications; 
(b) The decision boundaries get shifted owing to the crossbar non-idealities in hardware, resulting in the placement of certain data points into a different hypervolume leading to clean accuracy losses. However, due to gradient obfuscation owing to crossbar non-idealities, many data points remain restricted in their original hypervolumes even on adversarial perturbation ('violet arrow'). This results in lower adversarial losses and better adversarial robustness.}
\label{picadv}
\vspace{-2mm}
\end{figure}

Non-idealities in crossbars have so far been projected in a negative light since, they lead to degradation in clean accuracy when DNNs are mapped onto them. However, in this work, we show how non-idealities (or an increased value of NF for a crossbar) lead to robustness of DNNs against adversarial attacks. That is, we observe lower adversarial loss ($AL$) in crossbar-mapped DNNs with respect to the corresponding software implementation of the DNNs. \textit{We argue that non-idealities intrinsically lead to defense via gradient obfuscation against adversarial perturbations since gradient propagation, as discussed in Section~\ref{adv}, is crucial to initiate an adversarial attack}. The entire phenomenon has been pictorially summarized in \figurename{~\ref{picadv}}

\textbf{Quantifying the intuition in Fig. 2: }To support our gradient obfuscation argument, let us consider a DNN mapped onto crossbars as $f$. The net perturbation added to the input ($X$) in case of an adversarial attack is given by $\Delta=\epsilon~\times~sign(\nabla_{x} L(\theta,X,y_{true}))$ (refer to Section~\ref{adv}). Without loss of generality, we assume the loss function ($L$) from the hardware mapped DNN to be a function of the output current emerging out of a crossbar array ($I_{out}$), \textit{i.e.}, $L = f(I_{out})$.
Since DNNs are sufficiently linear owing to the Rectified Linear Unit (ReLU) activation, we can assume that, $L \approx I_{out}$. In the ideal scenario of crossbars with no non-idealities, $L \approx Iout_{ideal}$. This implies, 
\begin{equation} \label{eq:diffideal}
\Delta_{ideal}=\epsilon~\times~sign(\nabla_{x} (Iout_{ideal}))
\end{equation}
However, in the case of non-idealities being present in crossbar structures,
\begin{equation}
Iout_{non-ideal} = Iout_{ideal} - \gamma
\end{equation}
where, $\gamma$ denotes the deviation of the output current of the crossbar from its ideal value due to the inherent non-idealities. Hence, in the non-ideal scenario, we have:
\begin{equation} \label{eq:diffnonideal}
\Delta_{non-ideal}=\epsilon~\times~sign(\nabla_{x} (Iout_{ideal}~-~\gamma))
\end{equation}

From equation (\ref{eq:diffnonideal}), we find that there is a deviation in the adversarial perturbation from its ideal value owing to crossbar non-idealities. Hence, this explains the reason behind an altered displacement of data points in the high-dimensional subspace \textit{w.r.t.} the direction of displacement in case of a DNN evaluated on software (\figurename{~\ref{picadv}}(b)). 

\vspace{-3mm}
\section{Crossbar evaluation platform to incorporate non-idealities}
\label{sec:method}

\begin{figure}[t]
    \centering
    \includegraphics[width=0.8\linewidth]{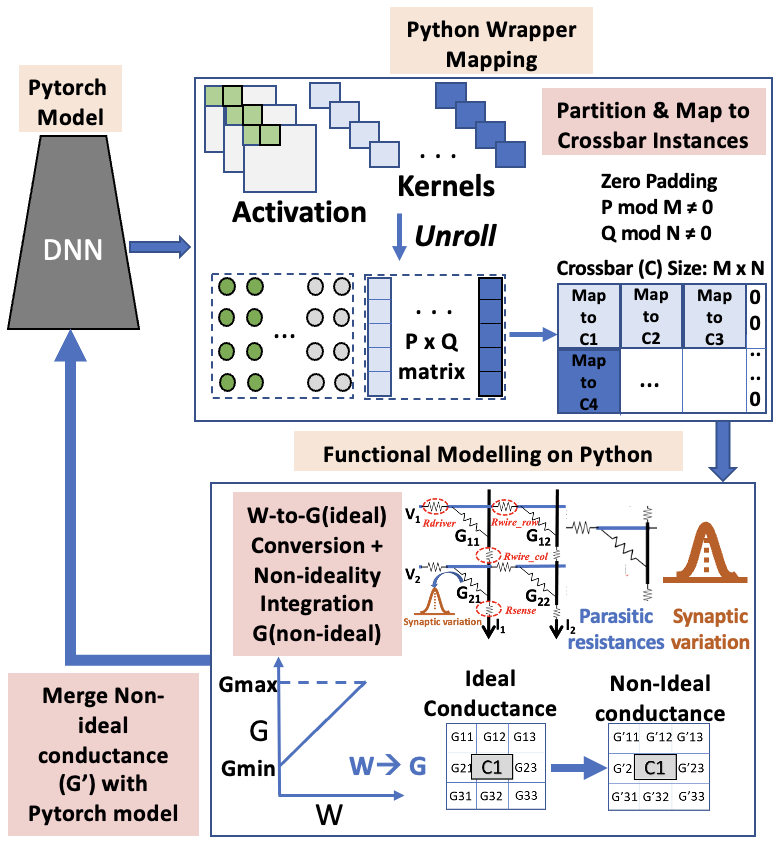}%
    \caption{Crossbar evaluation platform to incorporate the impact of non-idealities in DNN weights for different crossbar sizes and specifications}
    \label{format}
    \vspace{-4mm}
\end{figure}

In this work, we develop a framework in \textit{Pytorch} to map DNNs onto non-ideal memristive crossbars and investigate the cumulative impact of the circuit and device-level non-idealities on their adversarial robustness. Note, we do not carry out non-ideality aware retraining of DNNs to improve on the clean accuracy upon hardware-mapping.
\figurename{~\ref{format}} illustrates the overall simulation framework that is used for non-ideal crossbar evaluation. The entire platform is Python based to enable better integration between the software model and the simulation framework. The software DNN is implemented in Pytorch. Then, a Python wrapper is built that unrolls each and every convolutional operation in the DNN into MVM operations. The matrices obtained are then zero-padded (if the crossbar size and weight matrix size are not exact multiples of each other) and partitioned into different crossbar instances. The next stage of the platform converts the weights ($W$) to suitable conductance values ($G$) taking into account the synaptic device parameters, \textit{viz.} $G_{MIN}$, $G_{MAX}$ and bit-precision. Then, the non-idealities from different sources- circuit-level and device-level, are integrated. Here, resistive non-idealities are included using circuit laws (\textit{Kirchoff's laws} and \textit{Ohm's law}) and linear algebraic operations that are written in Python~\cite{rxnn}. The device variations are included with gaussian profiling. The non-ideal synaptic conductances ($G'$) are then integrated into the original Pytorch model to conduct inference. This framework enables us to follow a device agnostic approach for analyzing the impact of intrinsic crossbar non-idealities on DNNs during inference. 

In our experiments, unless specified otherwise, the memristive crossbars have synaptic devices with an ON/OFF ratio of 10 (\textit{i.e,} $R_{MIN} = 20 k\Omega$ and $R_{MAX} = 200 k\Omega$), and the resistive non-idealities as follows: $Rdriver = 1 k\Omega$, $Rwire\_row = 5 \Omega$, $Rwire\_col = 10 \Omega$ and $Rsense = 1 k\Omega$. Further, the stochasticity in the synaptic device conductance has been modelled as a Gaussian variation with $\sigma/\mu = 10\%$.

\vspace{-3mm}
\section{Experimental Methodology}

We conduct the crossbar robustness experiments on state-of-the-art VGG8 and VGG16 DNN architectures with benchmark datasets- CIFAR-10, CIFAR-100 and Tiny Imagenet. Note, Tiny Imagenet dataset is a subset of the Imagenet dataset having 100,000 training examples and 10,000 test examples from 200 different classes. After training the DNNs on software, we launch FGSM and PGD attacks by adding adversarial perturbations to the clean test inputs and record the adversarial loss ($AL$) in each case. We term these attacks as \textit{Attack-SW}. This forms our baseline results. Then, we map the DNN weights of the software models onto the non-ideal memristive crossbars using the crossbar evaluation platform as illustrated in \figurename{~\ref{format}}. We take into account crossbars with sizes 16$\times$16, 32$\times$32 and 64$\times$64 for our experiments. After obtaining the parameters for the crossbar-mapped DNN models, we measure the clean accuracies in each case. We then subject the models to FGSM or stronger PGD based adversarial attacks to assess the impact of the inherent non-idealities on adversarial robustness of the mapped networks. 

\textbf{Modes of adversarial attack: } We consider two modes of adversarial attack for the crossbar-mapped DNNs - (a) \textit{Software-inputs-on-hardware (SH) mode} where, the adversarial perturbations for each attack are created using the software DNN's loss function and then added to the clean input that yields the adversarial input. The generated adversaries are then fed to the crossbar-mapped DNN. This is an example of a Black-Box (BB) attack. (b) \textit{Hardware-inputs-on-hardware (HH) mode} where, the adversarial inputs are generated for each attack using the loss from the crossbar-mapped models. The HH perturbations will incorporate the effect of the intrinsic hardware non-idealities and can be regarded as White-Box (WB) attacks.

\vspace{-3mm}
\section{Results and Discussion}
\label{sec:results}

\subsection{Comparison of clean accuracies}

\begin{wrapfigure}{l}{0.25\textwidth}
 \centering
\includegraphics[width=0.25\textwidth]{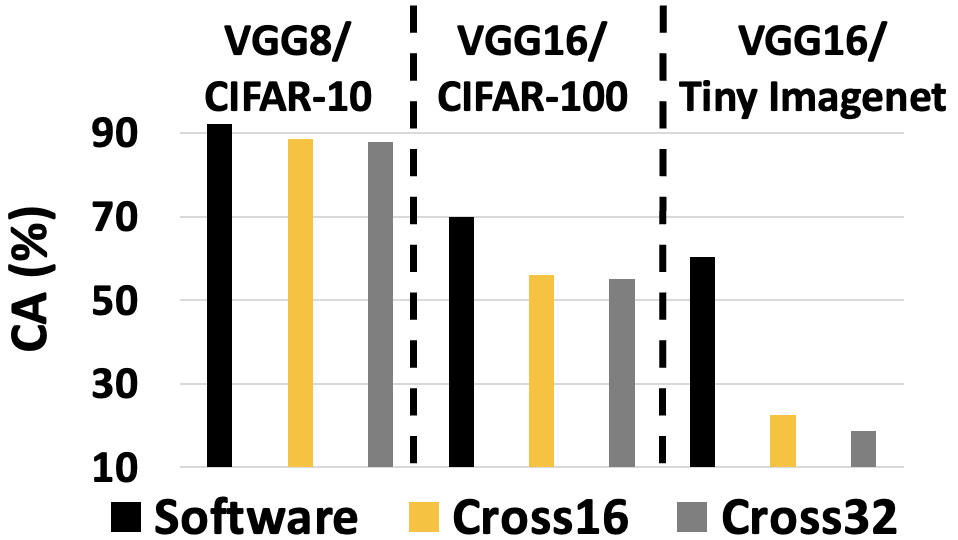}
  
\caption{Bar diagram showing clean accuracies of VGG8 and VGG16 networks on software and when mapped on crossbars of sizes 16x16 and 32x32}
\label{hist}
\end{wrapfigure}

\figurename{~\ref{hist}} presents a comparison of clean accuracies ($CA$) of the trained VGG8 and VGG16 networks when evaluated on software and after mapping on non-ideal crossbars of various dimensions. It can be observed that the $CA$ values drop post-mapping on the crossbars which is a direct implication of the inherent non-idealities causing errors in MVMs as discussed in Section \ref{nonideal}. Further, the $CA$ drops more for larger sized crossbars. Also, the drop in $CA$ post-mapping is more severe for networks trained with larger and complex datasets (\textit{e.g.,} Tiny Imagenet). In the subsequent subsections, we discuss the implications on adversarial robustness by inducing SH and HH attacks on the crossbar-mapped models.
\vspace{-3mm}
\subsection{Results with CIFAR-10 dataset}
\label{vgg8}

In \tablename{~\ref{tab:fgsm-cifar10-vgg8}}, it can be observed that $AL$ in case of an FGSM attack on the DNNs mapped onto crossbars of various dimensions (16$\times$16, 32$\times$32, 64$\times$64) are lesser than that of software baseline (i.e., Attack-SW). For different attack strengths quantified by $\epsilon$ values, $AL$ values in case of Attack-SW is significantly greater than SH or HH attacks ($>10-15\%$). In other words, non-idealities provide robustness against adversarial inputs by preventing the adversarial accuracies from degrading as severe as the clean accuracies.   

PGD attack, being a multi-step variant of FGSM attack, is much stronger and yields significantly higher adversarial losses in DNNs than FGSM attacks. 
Similar to the case of FGSM attacks, non-idealities in crossbars are seen to provide robustness to the mapped DNNs against adversarial inputs as shown in \tablename{~\ref{tab:pgd-cifar10-vgg8}}. Interestingly, we also find that larger crossbar sizes provide greater robustness against adversarial attacks (characterized by lower values of AL for the same value of $\epsilon$) than the smaller ones. This is because larger crossbars involve greater number of parasitic components (non-idealities), thereby imparting more robustness. This effect is more prominent in \tablename{~\ref{tab:pgd-cifar10-vgg8}} for stronger PGD attacks, where the crossbar size of 64$\times$64 provides the best robustness. 

\begin{table}[t]
\caption{Table showing the variation of $ALs$ with $\epsilon$ for FGSM attack on VGG8 network with CIFAR-10 dataset}
\label{tab:fgsm-cifar10-vgg8}
\resizebox{0.49\textwidth}{!}{%
\begin{tabular}{ccccccc}
\hline
\multicolumn{1}{c}{\textbf{$\epsilon$}} &
  
  \multicolumn{1}{c}{\textbf{0.05}} &
  \multicolumn{1}{c}{\textbf{0.1}} &
  \multicolumn{1}{c}{\textbf{0.15}} &
  \multicolumn{1}{c}{\textbf{0.2}} &
  \multicolumn{1}{c}{\textbf{0.25}} &
  \multicolumn{1}{c}{\textbf{0.3}} \\ \hline
\textbf{Attack-SW}  & 49.83 & 58.51 & 63.93 & 67.18 & 69.57 & 71.35 \\ \hline
\multicolumn{7}{c}{\textbf{Cross16}}                                 \\ \hline
\textbf{SH}         & 39.81 & 48.55 & 53.99 & 57.31 & 59.47 & 61.55 \\
\textbf{HH}         & 39.06 & 47.38 & 52.42 & 55.68 & 57.77 & 59.47 \\ \hline
\multicolumn{7}{c}{\textbf{Cross32}}                                 \\ \hline
\textbf{SH}         & 39.66 & 48.04 & 53.66 & 57.05 & 59.46 & 61.15 \\
\textbf{HH}         & 39    & 47.14 & 52.28 & 55.32 & 57.26 & 58.94 \\ \hline
\multicolumn{7}{c}{\textbf{Cross64}}                                 \\ \hline
\textbf{SH}         & 39.63 & 47.57 & 52.84 & 56.34 & 58.76 & 60.44 \\ \hline
\multicolumn{1}{c}{\textbf{HH}} &
 
  \multicolumn{1}{c}{40.08} &
  \multicolumn{1}{c}{47.83} &
  \multicolumn{1}{c}{52.61} &
  \multicolumn{1}{c}{55.87} &
  \multicolumn{1}{c}{58.12} &
  \multicolumn{1}{c}{59.81} \\ \hline
\end{tabular}%
}
\vspace{-3mm}
\end{table}

\begin{table}[t]
\caption{Table showing the variation of $ALs$ with $\epsilon$ for PGD attack on VGG8 network with CIFAR-10 dataset}
\label{tab:pgd-cifar10-vgg8}
\resizebox{0.49\textwidth}{!}{%
\begin{tabular}{cccccc}
\hline
\multicolumn{1}{c}{\textbf{$\epsilon$}} &
  \multicolumn{1}{c}{\textbf{2/255}} &
  \multicolumn{1}{c}{\textbf{4/255}} &
  \multicolumn{1}{c}{\textbf{8/255}} &
  \multicolumn{1}{c}{\textbf{16/255}} &
  \multicolumn{1}{c}{\textbf{32/255}} \\ \hline
\textbf{Attack-SW} & 83.66 & 83.75 & 83.97 & 84.87 & 86.88 \\ \hline
\multicolumn{6}{c}{\textbf{Cross16}}                     \\ \hline
\textbf{SH}        & 74.79 & 74.91 & 75.29 & 76.14 & 78.41 \\
\textbf{HH}        & 71.78 & 71.89 & 71.97 & 72.52 & 73.92 \\ \hline
\multicolumn{6}{c}{\textbf{Cross32}}                     \\ \hline
\textbf{SH}        & 73.48 & 73.58 & 74.1  & 74.9  & 77.54 \\
\textbf{HH}        & 71.13 & 71.22 & 71.55 & 71.97 & 73.36 \\ \hline
\multicolumn{6}{c}{\textbf{Cross64}}                     \\ \hline
\textbf{SH}        & 70.28 & 70.29 & 70.58 & 71.03 & 72.4  \\ \hline
\multicolumn{1}{c}{\textbf{HH}} &
  \multicolumn{1}{c}{67.92} &
  \multicolumn{1}{c}{67.14} &
  \multicolumn{1}{c}{67.38} &
  \multicolumn{1}{c}{68.73} &
  \multicolumn{1}{c}{71.06} \\ \hline
\end{tabular}%
}
\vspace{-4mm}
\end{table}

\subsection{Results with CIFAR-100 dataset}

\begin{figure}[t]
    \centering
    \includegraphics[width=0.85\linewidth]{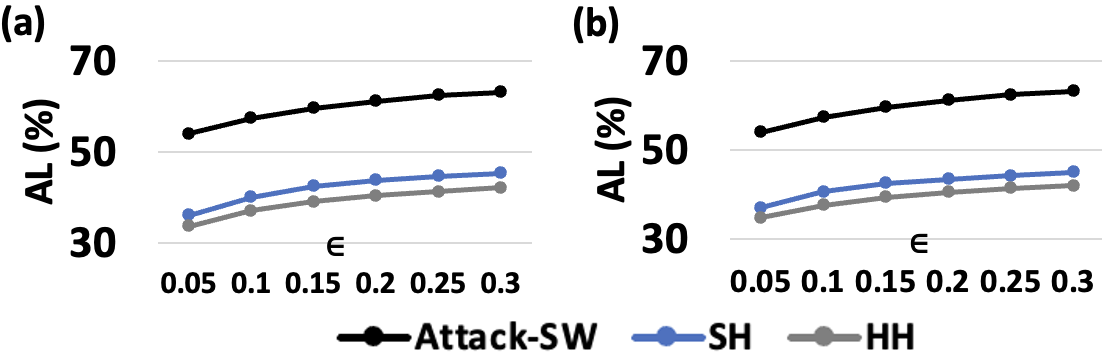}%
    %\label{}
    %\captionsetup{justification=centering}
    \caption{(a)-(b) A plot between AL and $\epsilon$ for Attack-SW, SH and HH attacks (FGSM) on VGG16 network with CIFAR-100 dataset for crossbar sizes 16$\times$16 and 32$\times$32 respectively}
    \label{vgg16-fgsm}
\end{figure}

\begin{figure}[t]
    \centering
    \includegraphics[width=0.95\linewidth]{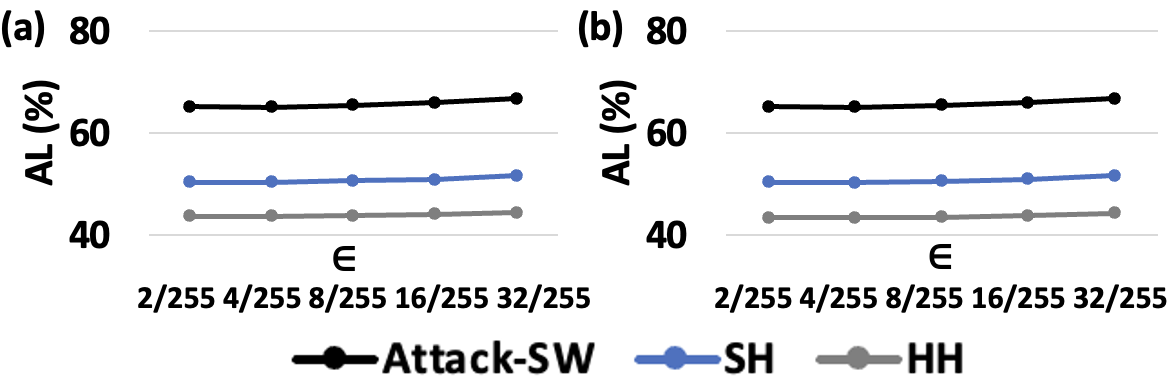}%
    %\label{}
    %\captionsetup{justification=centering}
    \caption{(a)-(b) A plot between AL and $\epsilon$ for Attack-SW, SH and HH attacks (PGD) on VGG16 network with CIFAR-100 dataset for crossbar sizes 16$\times$16 and 32$\times$32 respectively}
    \label{vgg16-pgd}
    \vspace{-4mm}
\end{figure}

The results shown in \figurename{~\ref{vgg16-fgsm}} and \figurename{~\ref{vgg16-pgd}} for a VGG16 network on CIFAR-100 dataset are similar to previous observations with CIFAR-10. Crossbar non-idealities impart adversarial robustness to the crossbar-mapped VGG16 network ($>10-20\%$) against both FGSM and PGD-based attacks. With CIFAR-100 dataset, we clearly observe that DNN shows much lower $AL$ values with PGD attack for HH mode than SH mode (than what was observed with CIFAR-10). Quantitatively, there is $\sim7\%$ (and $\sim4\%$) greater robustness in  HH attack \textit{w.r.t} SH attack with CIFAR-100 (and CIFAR-10) dataset, respectively. Furthermore, in CIFAR-100, with PGD attacks, $AL$ values tend to get saturated and do not increase upon increasing the perturbation strength $\epsilon$ for the crossbar-mapped models. Hence, the crossbar-mapped models show better resilience against increasing attack strengths for the stronger PGD attacks. 

\subsection{Results with Tiny Imagenet dataset}

\begin{wrapfigure}{l}{0.20\textwidth}
 \centering
\includegraphics[width=0.19\textwidth]{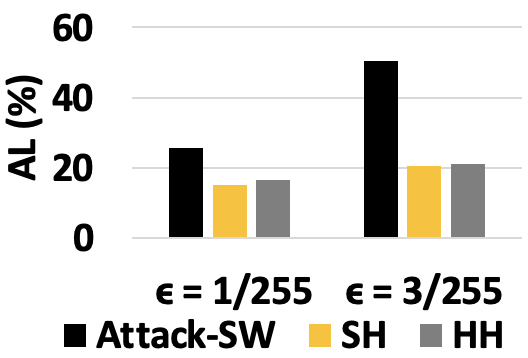}
  
\caption{Bar-diagram showing ALs for Attack-SW, SH and HH attacks (FGSM) on VGG16 network with Tiny Imagenet for crossbar size of 16$\times$16 with respect to $\epsilon~=~1/255,~3/255$}
\label{tinyimagenet}
\end{wrapfigure}
\figurename{~\ref{tinyimagenet}} shows the $AL$ for a VGG16 network on Tiny Imagenet dataset when subjected to FGSM attack with $\epsilon~=~1/255,~3/255$. Similar to previous results, we find that crossbar non-idealities confer advantage to DNNs in providing adversarial resilience in case of both SH and HH attacks. In this scenario, we find a significant $\sim11-37\%$ improvement in adversarial robustness for the DNN mapped onto 16$\times$16 crossbars with respect to the software baseline. 
\vspace{-4mm}
\subsection{Effect of crossbar parameters on adversarial robustness}

\subsubsection{Effect of $R_{MIN}$}

\begin{figure}[t]
    \centering
    \includegraphics[width=\linewidth]{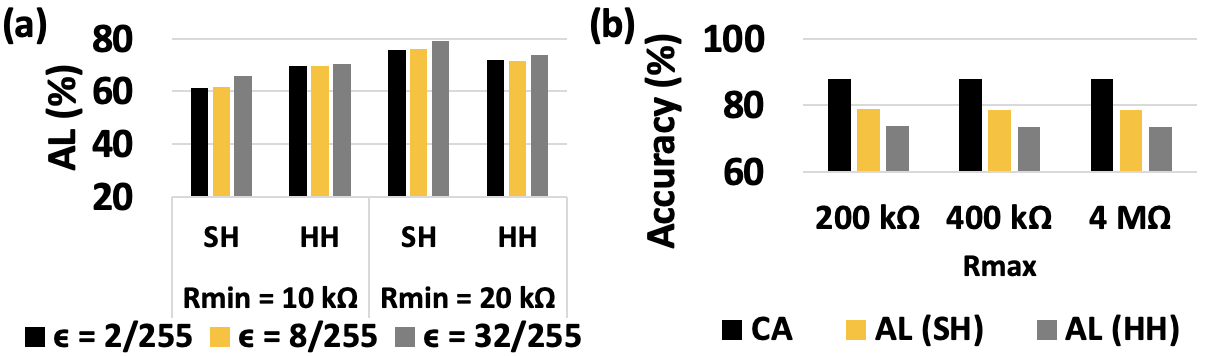}%
    %\label{}
    %\captionsetup{justification=centering}
    \caption{(a) Bar-diagram showing ALs in case of SH and HH attacks (PGD) for VGG8 network with CIFAR-10 dataset mapped on 32$\times$32 crossbars for two different values of $R_{MIN}$ at constant $R_{MAX}/R_{MIN}$ ratio; (b) Bar-diagram showing CAs and ALs (for PGD-based SH and HH attacks at $\epsilon = 32/255$) for a VGG8 network with CIFAR-10 dataset mapped on 32$\times$32 crossbars}
    \label{rmin}
\end{figure}

The effective resistance of a crossbar array is the parallel combination of resistances along its rows and columns. A smaller value of $R_{MIN}$ reduces the effective resistance of the crossbar and increases the value of NF for the crossbar~\cite{geniex}. We have already argued that an increased value of NF improves the adversarial robustness of crossbars. So, on decreasing $R_{MIN}$ to 10 $k\Omega$ (maintaining a constant $R_{MAX}/R_{MIN}$ ratio of 10), we find that ALs (for a PGD attack) in case of smaller $R_{MIN}$ are lower than the corresponding ALs for a larger $R_{MIN}$ as shown in \figurename{~\ref{rmin}} (a). 

\subsubsection{Effect of $R_{MAX}$ at constant $R_{MIN}$}
In \figurename{~\ref{rmin}} (b), we keep $R_{MIN}$ constant at 20 $k\Omega$ and increase $R_{MAX}/R_{MIN}$ ratio by increasing the value of $R_{MAX}$. We find that even increasing $R_{MAX}$ by a factor of 200 results in no added advantage in terms of adversarial robustness for SH or HH attacks. This is because the effective resistance of a crossbar array is the parallel combination of resistances along its rows and columns and hence, is largely determined by the value of $R_{MIN}$, which is kept constant. Hence, there is a greater impact of $R_{MIN}$ on adversarial robustness than $R_{MAX}$.

\subsection{Combining hardware mapping with efficiency-driven optimization techniques}

\begin{figure}[t]
    \centering
    \includegraphics[width=0.85\linewidth]{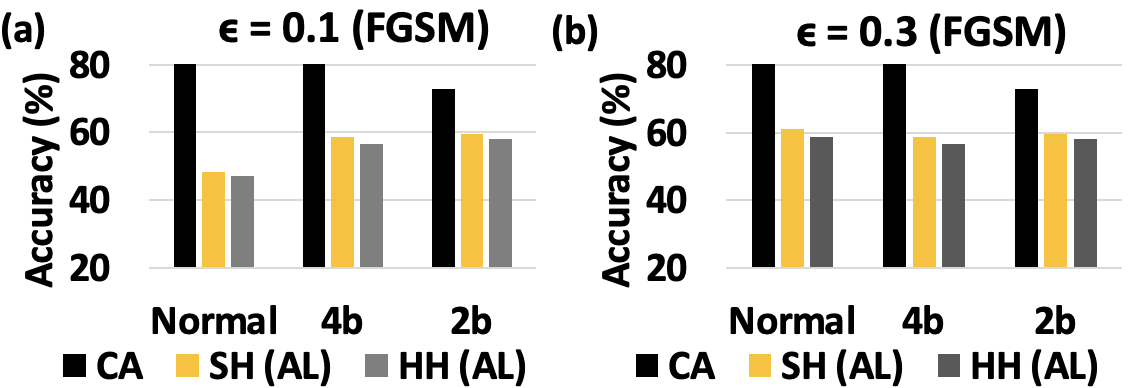}%
    %\label{}
    %\captionsetup{justification=centering}
    \caption{Bar-diagram showing CAs and ALs (for FGSM-based SH and HH attacks) for a VGG8 network mapped on 32$\times$32 crossbars using CIFAR-10 dataset for different bit-discretizations of input pixels (4-bit and 2-bit) with - (a) $\epsilon = 0.1$; (b) $\epsilon = 0.3$}
    \label{disx-xbar}
    \vspace{-4mm}
\end{figure}

\begin{table}[t]
\caption{Table showing the variation of $ALs$ with $\epsilon$ for FGSM attack on full-precision (F) and QUANOS-based hybrid-precision (Q) VGG8 networks with CIFAR-10 dataset mapped onto 16$\times$16 crossbars.}
\label{tab:w_quanos}
\resizebox{0.5\textwidth}{!}{%
\begin{tabular}{ccccccc}
\hline
\textbf{$\epsilon$} & \textbf{0.05} & \textbf{0.1} & \textbf{0.15} & \textbf{0.2} & \textbf{0.25} & \textbf{0.3} \\ \hline
\textbf{Attack-SW (F)} & 49.83 & 58.51 & 63.93 & 67.18 & 69.57 & 71.35 \\ 
\textbf{Attack-SW (Q)} & 48.2  & 54.36 & 59.15 & 62.52 & 65.5  & 67.94 \\ \hline
\textbf{SH (F)}        & 39.81 & 48.55 & 53.99 & 57.31 & 59.47 & 61.55 \\ 
\textbf{SH (Q)}        & 21.29 & 23.02 & 23.78 & 25.2  & 26    & 26.63 \\ \hline
\textbf{HH (F)}        & 39.06 & 47.38 & 52.42 & 55.68 & 57.77 & 59.47 \\
\textbf{HH (Q)}        & 23.85 & 26.51 & 27.78 & 28.72 & 29.29 & 30    \\ \hline
\end{tabular}%
}
\end{table}

\subsubsection{Combined effect of input pixel discretization and crossbar non-idealities}

In \cite{pixeld}, the authors show that input pixel discretization from 256 or 8-bit level to 4-bit, 2-bit improves adversarial resilience of software DNNs. Here, we unleash FGSM attack on the VGG8 network mapped onto 32$\times$32 crossbars with input image pixels of the CIFAR-10 test dataset discretized to 4-bits (or 16 levels) and 2-bits (4 levels). The results are shown in \figurename{~\ref{disx-xbar}}. Interestingly, we find that with pixel discretization, ALs on crossbar mapped DNN for both SH and HH attacks attain a fixed value and do not vary on increasing $\epsilon$ from 0.1 to 0.3. This implies that input pixel discretization does not necessarily help in resiliency when attacking hardware mapped DNNs. For lower $\epsilon$ ($\epsilon = 0.1$), greater adversarial robustness is observed without pixel discretization. At higher value of $\epsilon$ ($\epsilon = 0.3$), the combined effect of 4-bit pixel discretization and crossbar non-idealities outperforms the rest in terms of adversarial robustness. Furthermore, 2-bit pixel discretization not only reduces the clean accuracy to $72.89\%$ but also imparts marginally lesser adversarial robustness than 4-bit pixel discretization - $<0.8\%$ for SH attack and $<1.48\%$ for HH attack. 

\subsubsection{Combined effect of QUANOS and crossbar non-idealities}

%We implement a DNN model trained using QUANOS~\cite{quanos}, by performing layer-specific hybrid quantization of its weights and activations based on the \textit{Adversarial Noise Sensitivity} (ANS) metric that make DNNs robust against adversarial perturbations. 
QUANOS allows us to
train a model with reduced bit-width at each layer based on the \textit{Adversarial Noise Sensitivity} (ANS) metric that make DNNs robust against adversarial perturbations. This translates to an
optimal quantized network with inference as well as training
energy savings on hardware. In \tablename{~\ref{tab:w_quanos}}, we show $AL$ values for FGSM attack on a hybrid-precision VGG8 network with CIFAR-10 dataset trained using QUANOS, mapped on 16$\times$16 crossbars, and compare the same with a standard full-precision VGG8 network. QUANOS, in itself, is a software defense strategy that makes DNNs resilient to adversarial attacks. When combined with crossbar non-idealities, the benefits are expected to increase. We find that the hybrid-precision model (Q) on hardware shows $\sim15-30\%$ lower $AL$ than the corresponding standard full-precision model (F) mapped on hardware. 
\vspace{-4mm}

\subsection{Comparison with standalone efficiency-driven defenses}

\begin{figure}[t]
    \centering
    \includegraphics[width=\linewidth]{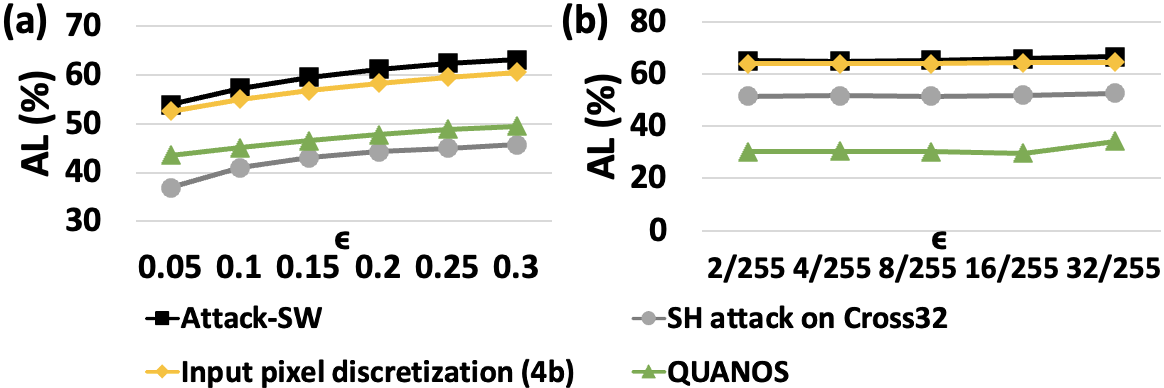}%
    %\label{}
    %\captionsetup{justification=centering}
    \caption{Comparison of our proposed method with other state-of-the-art adversarial defenses using VGG16 network and CIFAR-100 dataset during (a) FGSM and (b) PGD attacks.}
    \label{compare}
    \vspace{-4mm}
\end{figure}

We compare the performance of non-ideality-driven adversarial robustness in crossbars against state-of-the-art software-based adversarial techniques described in~\cite{pixeld, quanos}. Note, \cite{pixeld, quanos} use efficiency driven transformations (that \textit{implicitly} translate to hardware benefits) such as, quantization to improve resilience. In contrast, our work utilizes \textit{explicit} hardware variations to improve robustness. We aim to compare the robustness obtained from implicit and explicit hardware techniques. We observe that for single-step FGSM attack on a VGG16 network with CIFAR-100 dataset mapped on 32$\times$32 crossbars, adversarial robustness due to crossbar non-idealities (SH attack results) outperforms all other techniques (\figurename{~\ref{compare}(a)}). For multi-step PGD attack, SH attack ranks second ((\figurename{~\ref{compare}(b)}). With respect to 4-bit (4b) \textit{pixel discretization} of input data~\cite{pixeld}, non-idealities in crossbars impart $\sim15\%$ greater adversarial robustness in case of FGSM attack and $\sim12\%$ greater adversarial robustness in case of PGD attack. On the other hand, in case of FGSM attack, crossbar non-idealities impart $\sim4\%$ greater adversarial robustness than QUANOS~\cite{quanos}, while for PGD attack, QUANOS outperforms by $\sim18-22\%$. 

\vspace{-5mm}

\section{Conclusion}
\label{sec:conclusion}

In this work, we perform a comprehensive analysis to show how crossbar non-idealities can be harnessed for adversarial robustness. This work brings in a new standpoint that does not devalue the importance of non-idealities or parasitics present in crossbar systems. We develop a framework based on \textit{Pytorch} that maps state-of-the-art DNNs (VGG8 and VGG16 networks) onto resistive crossbar arrays and evaluates them with benchmark datasets (CIFAR-10, CIFAR-100 and Tiny Imagenet). We show that circuit-level non-idealities (\textit{e.g.}, interconnect parasitics) and device-level non-idealities provide robustness to the mapped DNNs against adversarial attacks, such as FGSM and PGD attacks. This is reflected by lower accuracy degradations during adversarial attacks in case of DNNs mapped on crossbars than that of software-based DNNs ($>10-15\%$). We further compare the performance of our non-ideality driven approach with other state-of-the-art software-based adversarial defense techniques on CIFAR-100 dataset. We find that our approach performs significantly well in terms of reducing adversarial losses during FGSM or PGD attacks. Our comprehensive analysis and encouraging results establish the idea of rethinking analog crossbar computing for adversarial security in addition to energy efficiency.

\section*{Acknowledgement}
This work was supported in part by C-BRIC, Center for Brain-inspired Computing, a JUMP center sponsored by DARPA and SRC, the National Science Foundation (Grant\#1947826), the Technology Innovation Institute, Abu Dhabi and the Amazon Research Award.
\vspace{-3mm}

\def\bibfont{\footnotesize}
\printbibliography
%\printbibitembibliography

\end{document}